\documentclass[aps,prd,superscriptaddress,preprint,tightenlines,nofootinbib]{revtex4}

\usepackage{graphicx} 
\usepackage{dcolumn}  
\usepackage{bm}       

\begin{document}

\preprint{CLNS 08/2022}       
\preprint{CLEO 08-06}         

\title{First Observation of the Decay $D_s^+ \to p \bar{n}$}


\author{S.~B.~Athar}
\author{R.~Patel}
\author{J.~Yelton}
\affiliation{University of Florida, Gainesville, Florida 32611, USA}
\author{P.~Rubin}
\affiliation{George Mason University, Fairfax, Virginia 22030, USA}
\author{B.~I.~Eisenstein}
\author{I.~Karliner}
\author{S.~Mehrabyan}
\author{N.~Lowrey}
\author{M.~Selen}
\author{E.~J.~White}
\author{J.~Wiss}
\affiliation{University of Illinois, Urbana-Champaign, Illinois 61801, USA}
% ...
\collaboration{CLEO Collaboration}
\noaffiliation


\date{March 10, 2008}

\begin{abstract} 

Using $e^+e^-\to D_s^{*-}D_s^+$ data collected near the peak $D_s$
production energy,  
$E_{cm}=4170$ MeV, with the CLEO-c detector, we present the 
first observation of the decay $D_s^+\to p \bar{n}$. We measure a 
branching fraction ${\cal{B}}(D_s^+\to p \bar{n}) = 
(1.30\pm 0.36 ^{+0.12}_{-0.16}) \times 10^{-3}$. This is the first observation
of a charmed meson decaying into a baryon-antibaryon final state.
\end{abstract}

\pacs{13.25.Ft,14.40.Lb}
\maketitle

Of the three ground state charmed mesons, only the $D_s^+$ is massive enough to decay
to a baryon-antibaryon pair. Even before the discovery of the $D_s^+$, a search for
the decay $D_s^+ \to p \bar{n}$ was suggested \cite{Pham} 
as a ``smoking gun'' for decays proceeding via annihilation through a virtual $W^+$,
and a prediction was made 
that the branching fraction would be $\approx 1\%$ if the annihilation
mechanism dominated $D_s^+$ decays. In the intervening period it has become clear that
the annihilation diagram contributes to, 
but does not dominate, $D_s^+$ decays, and has been 
studied in purely leptonic decays such as $D_s^+ \to \mu\nu$ \cite{munu} and
$D_s^+ \to \tau\nu$ \cite{taunu}. 
However, although
the theoretical study of $D_s^+ \to p \bar{n}$ is complicated by
final state interactions, it still has a unique role to play in the 
understanding of charmed meson decays.

Finding decay modes that include an (anti-)neutron is particularly challenging. 
CLEO-c is the first detector to have a large dataset ($\approx 325\ \rm{pb}^{-1})$
of $e^+e^-$ annihilation 
events taken at a center-of-mass energy of around 4170 MeV. At this energy, there
is a substantial cross-section ($\approx 1\ $nb)
for the reaction $e^+e^- \to D_s^{*-}D_s^+$ \cite{DCH}. Using the 
knowledge of center-of-mass energy and momentum, we use a missing mass method to
find the (anti-)neutron and so do not depend upon its interaction in the detector. 
Therefore, in this paper, mention of a particular decay mode implies 
the use of the
charge conjugate decay mode also.

The CLEO-c detector \cite{CLEO} is designed to measure the momenta of charged
particles which curve in a 1.0 T solenoidal magnetic field,
and identify them using specific energy loss ($dE/dx$) and Cherenkov
imaging (RICH). Photons are detected, and their energy measured, using a 
CsI calorimeter. Our analysis procedure has much in common with that used 
in the measurement of ${\cal{B}}(D_s^+\to \mu\nu)$ \cite{munu}. Here, 
we fully reconstruct one $D_s^-$
as a ``tag,''
reconstruct a transition photon from a $D_s^*$ decay, and identify and measure the 
momentum of a proton. 
We can then reconstruct the missing mass of the event and look for a peak
at the anti-neutron mass.

The $D_s^-$ tags are found in the eight modes: $K^+K^-\pi^-$, $K^0_sK^-$, $\eta\pi^-$,
$\eta^{\prime}\pi^-$, $\phi\rho^-$, $\pi^-\pi^+\pi^-$, $K^{*-}K^{*0}$ and $\eta\rho^-$.
Track selection, particle identification and definition of resonances are similar to
those in our previous publication \cite{munu}, with one important exception; each of the 
charged tracks in the $D_s^-$ tag is required to have a $dE/dx$ measurement more than three 
standard deviations, $\sigma$, 
away from that expected for a proton. All the $D_s^-$ tags are required to have momentum
consistent with coming from the two-body production $D_s^*D_s$. 
Figure~1 shows the measured mass of the tag candidate 
minus the nominal $D_s^-$ mass,
divided by the resolution of the candidate tag's mode.
The fit shown is a unit Gaussian centered at zero. 
Those candidates within 2.5 of the peak 
are kept as $D_s^-$ tags. According to the fit they comprise 27700 
real $D_s^-$ mesons and 64900 background combinations. Those combinations
in the regions $\pm(3.5-6.0)$ in this plot will be used as a check on the combinatorial 
background to our final signal. We kinematically constrain the mass of the 
tags in the signal region to the
known mass of the $D_s^-$. To be consistent, each sideband tag is 
kinematically constrained to the center of its sideband.

\begin{figure}
\includegraphics*[width=3.75in]{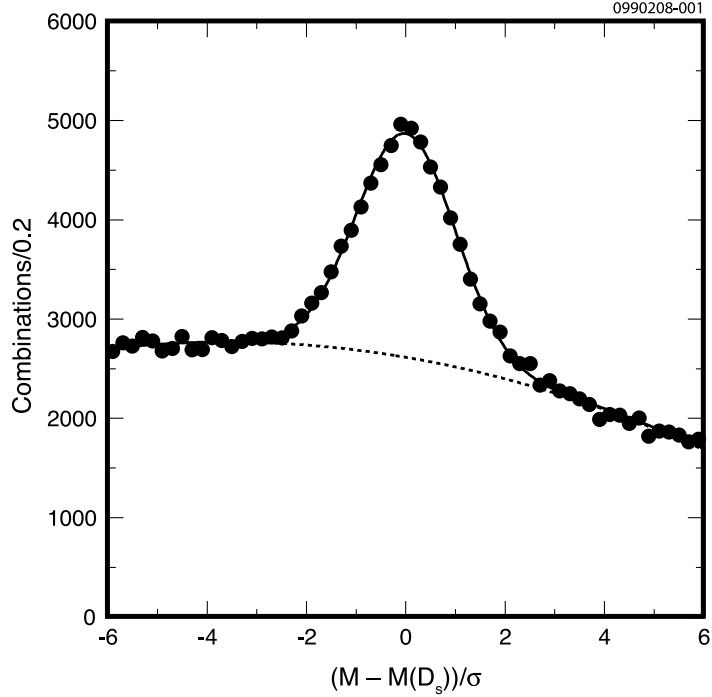}
\caption{The reconstructed mass minus the known $D_s$ mass, divided
by the detector resolution, for all eight modes of $D_s$ tags reconstructed.
The fit shown is a unit Gaussian centered at zero, together with a second order
polynomial background function.}
\label{fig:Ds}
\end{figure}

We now add a $\gamma$ that satisfies our shower 
shape requirements to the $D_s^-$ tag. 
Using the four-momenta of the $\gamma$ candidate, $p(\gamma)$, 
and the $D_s^-$ tag, we calculate
the four-momentum of the $D_s^+$ using the equation $p(D_s^+)=p_{beam}-p(D_s^-)-p(\gamma)$.
The four-momentum of the beam, $p_{beam}$ takes into account the small crossing angle of the 
CESR beams. The missing mass squared (${\rm MM^2}$)
distribution for good tag events is shown in Fig.~2. It shows
a peak at $M^2(D_s^{+})$ corresponding to $D_s^*D_s$ production. 
This is fit to a signal shape of a Crystal Ball function \cite{CB} with fixed tail parameters
derived from Monte Carlo simulation, 
together with a fifth order polynomial background
function. 
We select those events with ${\rm MM}^2$ values between 3.779 and 3.976 $\rm GeV^2$. 
This is a
loose requirement, with most of the loss in efficiency due to 
initial state radiation of the
beam, which smears the ${\rm MM}^2$ to artificially high values. 
According to the fit, the yield of $D_s^- \gamma D_s^+$ candidates in 
this range is $16955$ above a background of 63170.
This yield will be the denominator in our final branching fraction calculation.

\begin{figure}
\includegraphics*[width=3.75in]{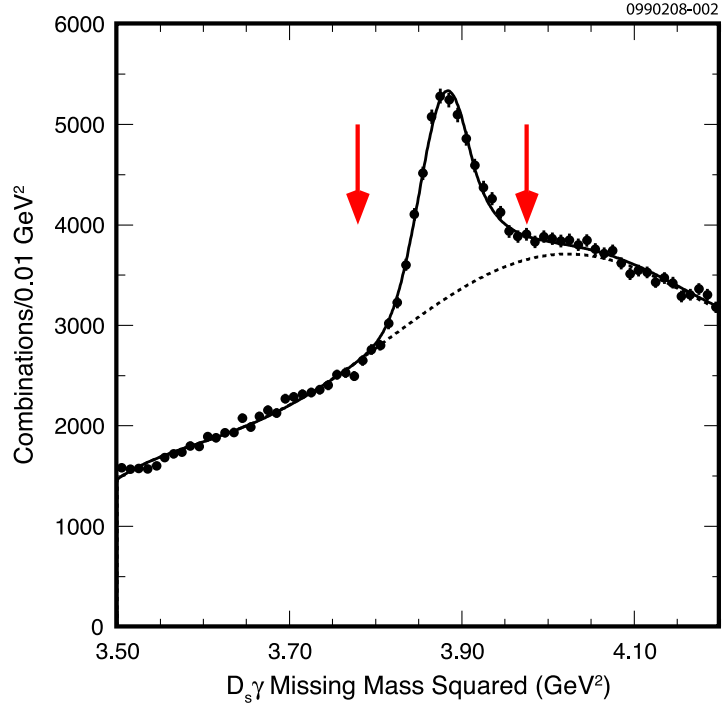}
\caption{The missing mass squared 
from events with a reconstructed $\gamma$ and $D_s(tag)$.
The fit is to a signal shape of a Crystal Ball function \cite{CB} with fixed tail parameters, 
together with a fifth order polynomial background
function. 
}
\label{fig:Dss}
\end{figure}

We next select our proton candidate. Monte Carlo simulation shows that all 
protons from this decay mode
will have momenta in the range 150-550 MeV/$c$. This is below the momentum 
range for the RICH detector
to identify protons, but well suited to identification by  $dE/dx$. 
We require that this measurement be within $3\sigma$ of that expected for a proton,
and greater than $3\sigma$ from that expected for a kaon or a pion. 
The overall proton efficiency is determined by Monte Carlo simulation to be 75\%,
with the efficiency lowest at the lower proton momenta.  

We may now calculate the missing 4-momentum in the event, 
equal to the expression $p_{beam}-p(D_s^-)-p(\gamma)-p_{proton}$, 
and can thus calculate the missing mass
of the event. However, we have further kinematic
constraints we can impose which allow us to improve the 
missing mass resolution and reject combinatorial 
background. We do not know $a\ priori$ if the photon is due to the 
transition $D_s^{*-}\to D_s^-(tag)\gamma$, or 
$D_s^{*+}\to D_s^+(signal)\gamma$. We perform kinematic fits 
with each assumption, and choose between 
the two based on the $\chi^2$ values of the two fits. 
First, we add the photon to the $D_s^{-}$ tag to form a 
$D_s^{*-}$ candidate, and 
constrain the momentum of this $D_s^{*-}$ candidate to that calculated from the 
two body production 
$e^+e^- \to D_s^{*-}D_s^+$. 
We then constrain the mass difference 
$M(D_s^{*-})-M(D_s^-)$ to 
its nominal value. Alternatively, we constrain the $D_s^-$ tag itself to the 
momentum calculated assuming
the two-body production $e^+e^- \to D_s{^-}D_s^{*+}$, then 
combine the proton with the missing mass of the event to make a $D_s$ signal 
candidate, add the photon, and constrain the $M(D_s^{*+}-D_s^+)$ mass difference. 
We choose the scheme with the lowest total
$\chi^2$ value; in Monte Carlo simulation we find that 
we assign the photon 
to the correct $D_s$ greater than 95\% of the time. 
The kinematic constraints on the detected
particles improve the resolution in missing mass by around a factor of two,
whichever $D_s$ the photon is combined with.
Furthermore, we can place cuts on the $\chi^2$ of the kinematic constraints to reject
combinatorial background. 
In the case of the 
momentum constraint we require $\chi^2 < 9$, 
and in the case of the mass-difference constraint
we require $\chi^2 <4$; with each constraint there is one degree of freedom. 
The requirement is looser for the
momentum constraint because initial state radiation produces a 
tail in the momentum distribution.

The transition photon in the event has an energy in the laboratory of 110-180 MeV. 
In this energy range there
is the possibility of background clusters passing 
all the requirements for being a photon.
Such background photons are particularly prevalent in 
events which contain anti-baryons as they 
frequently interact with the detector and give ``split-off'' clusters, 
often far from the impact point of the 
particle in the CsI calorimeter. 
Occasionally an event may survive all the above requirements while having more
than one photon candidate. If so, we select the
photon candidate that produces the lowest
combined $\chi^2$ in the kinematic fit. Background photons also influence the 
signal shape which we determine using Monte Carlo simulation. This shape is well described by a 
core Gaussian function of $\sigma \approx 4$ MeV centered at the neutron mass, 
together with a second, offset, Gaussian of width $\sigma \approx 38 $ MeV and containing
$\approx12\%$ of the signal. 
This second Gaussian is due to events where we have used
an incorrect photon candidate.

Figure~3 shows the missing mass distribution for the events after all requirements
and kinematic fitting, and contains thirteen events. These are the only events
in the missing-mass range 600-1100 MeV. The plot is well fit
using a likelihood fit to the signal shape described above, 
so we take our signal yield to be the $13.0\pm3.6$ events observed.
Repeating the analysis using sidebands to the $D_s^-$ as described above, gives
three events in the missing-mass range 600-1100 MeV, none of which are in the 
signal region of 900-980 MeV.
We divide this yield of 13 by the number of $D_s^+$ decays we have detected, and correct
for the efficiencies of requirements placed on the fit $\chi^2$ and proton reconstruction
and identification. This gives a branching fraction of $(1.30\pm 0.36)\times 10^{-3}$,
where the error shown is the statistical error in the signal yield only.

\begin{figure}
\includegraphics*[width=3.75in]{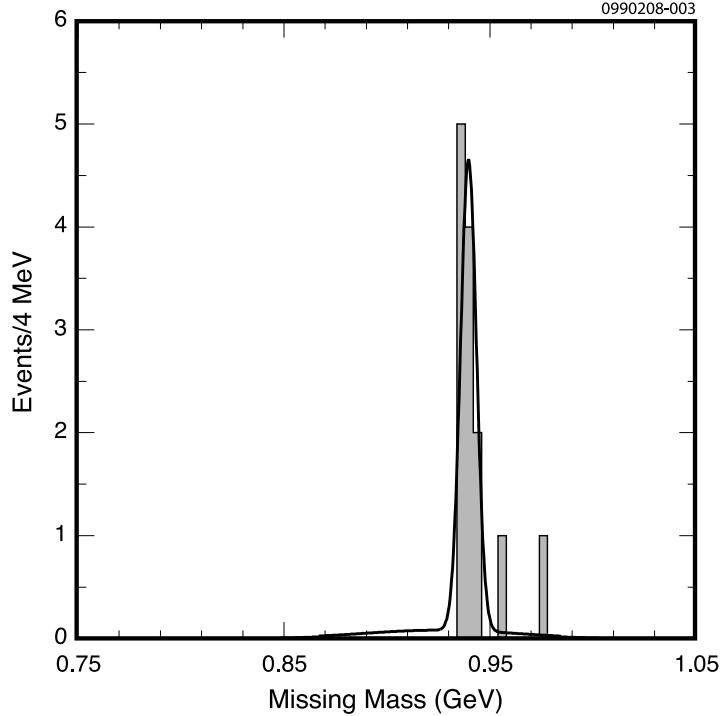}
\caption{The missing mass in the event after all requirements and 
kinematic fitting has been performed. The fit is described in the text.}
\label{fig:MM}
\end{figure}

We have performed many checks to ensure that our analysis
is not biased towards
obtaining events only in the signal region. To enter the final signal plot, an event
must have a proton candidate of a momentum which happens to be
well matched to the capabilities of the 
$dE/dx$ system, and thus background from incorrectly identified proton
candidates is negligible. 
This, in turn, means that background from other charm events is negligible. 
In order to check that the sidebands are a reasonable representation of 
combinatorial background, we generated a large sample of $uds$ continuum Monte Carlo 
events with final states of a proton, an anti-neutron, a
photon, and the decay products of a $D_s^-$. Very
few of these events passed all the kinematic requirements, and among those that did,
there were as many
anti-neutron candidates in the sideband plot as the signal plot. 

As a check on the efficiency of the selection criteria imposed on the quality of the
kinematic fit, we repeated the analysis
looking for the decay $D_s^+ \to K^+K^0$, where we consider the $K^0$ to be the missing-mass
analog of the antineutron. We reproduce very well the measurement of this branching fraction
found by more direct means \cite{Onyisi}, and from the comparison of the efficiencies 
of the kinematic constraint cuts in data and Monte Carlo, we find a systematic uncertainty
in the efficiency of $\pm$4\%. The uncertainty in the number of $D_s^+$, the denominator in our
branching fraction calculation, is systematically limited, and by looking at
the variation of the yield using a variety of different signal and background functions, 
estimated to be $\pm$5\%. 
One possible source of systematic uncertainty concerns the signal shape, 
the tail of which is dominated
by split-offs from antibaryon interactions which are one of the hardest
processes to reliably simulate. We conservatively assign a $\pm$6\% uncertainty to 
account for any mis-modelling of this process; this number corresponds to getting 
the number of events from these fake photons incorrect by $\pm$50\%.
Lastly, we note that although there is no evidence of any background in the final plot,
we cannot assume that it is strictly zero. As our best estimate of the background
is 0 events, which corresponds to an upper limit (1 $\sigma$) of 1.1 events,
we therefore introduce a -8.5\% uncertainty in the final branching fractions. 
Combining these systematic uncertainties
in quadrature, produces a total systematic uncertainty of $^{+9}_{-12}\%$ in the 
branching fraction, much smaller than the statistical error.

In conclusion, we report the first 
observation of 
the decay $D_s^+ \to p\bar{n}$ with 
a signal of 13 events and a background consistent with 
zero. 
We measure the branching fraction
${\cal{B}}(D_s^+ \to p \bar{n}) = (1.30\pm0.36^{+0.12}_{-0.16}) \times 10^{-3}$. 
This is the first observation of
a charm meson decaying into baryon-antibaryon pair. 
The two-body
decay observed here is the only one allowed kinematically. The actual
decay process is suspected to be related to annihilation, which is also responsible
for purely leptonic decays. Relating this baryonic decay rate to the leptonic rate 
should provide important clues as to how baryons are produced in hadronic interactions.

We gratefully acknowledge the effort of the CESR staff 
in providing us with excellent luminosity and running 
conditions. This work is supported by the A.P.~Sloan 
Foundation, the National Science Foundation,
the U.S. Department of Energy, the Natural Sciences and Engineering
Research Council of Canada, and the U.K. Science and Technology
Facilities Council.


\begin{thebibliography}{99}

\bibitem{Pham}
Xuan Yem Pham, Phys. Rev. Lett. {\bf 45}, 1663 (1980).


\bibitem{munu}
M.~Artuso {\it et~al.} (CLEO Collaboration), Phys. Rev. Lett {\bf 99}, 071802 (2007); 
T.~Pedlar {\it et~al.}, (CLEO Collaboration), Phys. Rev. { D}{\bf 76}, 072002 (2007).

\bibitem{taunu}
K.~M.~Ecklund {\it et~al.} (CLEO Collaboration), arXiv:0712.1175[hep-ex], submitted to Phys. Rev. Lett.

\bibitem{DCH}
D.Cronin-Hennessy {\it et al.} (CLEO Collaboration), arXiv:0801.3418[hep-ex], submitted
to Phys. Rev. D.

\bibitem{CLEO} 
Y.~Kubota {\it et~al.} (CLEO Collaboration), {Nucl. Instrum. Meth. A} \textbf{320}, {66} ({1992});
D.~Peterson {\it et~al.}, {Nucl.Instrum. and Meth. A} \textbf{478}, {142} 
({2002});
R.A.Briere {\it et al.} (CESR-c and CLEO-c Taskforces, CLEO-c Collaboration), 
Cornell University, LEPP Report No. CLNS 01/1742 (2001) (unpublished).

\bibitem{CB} P.~Rubin {\it et~al.} (CLEO Collaboration), Phys. Rev. {D} {\bf 73}, 112005 (1996). 

\bibitem{Onyisi}
J.~Alexander {\it et~al.} (CLEO Collaboration), arXiv:0801.0680[hep-ex], submitted to Phys. Rev. Lett.


\end{thebibliography}
\end{document}